# Impact of positive ion energy on carbon-surface production of negative ions in deuterium plasmas

https://doi.org/10.1088/1361-6463/ab34f1

**D. Kogut[1], R. Moussaoui[1], Ning Ning[1], J. B. Faure[1], J. M. Layet[1], T. Farley[2], J. Achard[4], A. Gicquel[4], G. Cartry[1]**

[1]*Aix-Marseille Université, CNRS, PIIM, UMR 7345, Centre Scientifique de Saint Jérôme, case 241, 13397 Marseille Cedex 20, France*
[2]*Department of Electrical Engineering and Electronics, The University of Liverpool, L69 3GJ, UK*
[4] *LSPM, CNRS-UPR 3407 Université Paris 13, 99 Avenue J. B. Clément, F-93430 Villetaneuse, France*

## Abstract

This work focuses on the production of negative-ions on graphite and diamond surfaces bombarded by positive ions in a low pressure (2 Pa) low power (20 W) capacitively coupled deuterium plasma. A sample is placed opposite a mass spectrometer and negatively biased so that surface produced negative ions can be self-extracted from the plasma and measured by the mass spectrometer. The ratio between negative-ion counts at mass spectrometer and positive ion current at sample surface defines a relative negative-ion yield. Changes in negative-ion production yields versus positive ion energy in the range 10-60 eV are analysed. While the negative-ion production yield is decreasing for diamond surfaces when increasing the positive ion impact energy, it is strongly increasing for graphite. This increase is attributed to the onset of the sputtering mechanisms between 20 and 40 eV which creates negative ions at rather low energy that are efficiently collected by the mass spectrometer. The same mechanism occurs for diamond but is mitigated by a strong decrease of the ionization probability due to defect creation and loss of diamond electronic properties.

Corresponding author: gilles.cartry@univ-amu.fr

# 1. Introduction

Negative-ions (NI) in low pressure plasmas are created in the plasma volume by dissociative attachment of electrons on molecules (volume production)[1] or on surfaces surrounding the plasma by bombardment of positive ions or hyper-thermal neutrals (surface production)[2,3,4,5,6]. The former is of primary importance in plasmas with highly electronegative gases such as those used in microelectronic processes[7,8,9] or for innovative applications such as space propulsion[10]. The latter is particularly strong when low-work function materials such as cesium are put in contact with the plasma[11,12] and this effect is employed in many types of negative-ion sources[13-20].

In magnetically confined fusion devices (tokamaks), the Neutral Beam Injector (NBI) accelerates NI to generate a fast neutral beam through interaction with a stripping gas target. Such beam is injected in the plasma to get heating and current drive. The huge dimensions of the ITER device and its successor DEMO compared to the present days tokamaks require neutral beam energies in the range of 1 MeV[21-25] where the neutralization of positive ions becomes very inefficient. Indeed, the yield of neutralization of the positive ions by a $D_2$ gas reaches zero above 100 keV, while its value is around 55% at 1 MeV for NI[26]. Therefore, there is a great research effort dedicated to the development of a high current NI source (40A $D^-$ beam for ITER)[19,27]. In these sources, extracted negative-ions are formed on the cesium-covered plasma grid which marks the transition between the plasma source and the accelerator region. The plasma grid is biased few volts below the local plasma potential leading to positive ion flux on the plasma grid dominated by low energy ions (on the order of few eV) with a tail up to tens of eV[28]. Hydrogen atoms also impact on the grid with energy distribution from tens of meV (thermalized atoms) to few eV (atoms resulting from $H_2$ dissociation before thermalization) or even tens of eV (atoms resulting from charge exchange collisions). Nonetheless, the average energy of the atomic flux on the plasma grid remains low, around 0.3 eV[29]. These sources use cesium injection inside the plasma in order to increase strongly the NI extracted current. However, as the use of cesium complicates noticeably the neutral beam injection device, there is a demand for the development of cesium-free NI sources in $H_2/D_2$ plasmas. Within this framework, we are studying NI surface production in cesium-free low-pressure $H_2/D_2$ plasmas.

Carbon materials, graphite and diamond, have been chosen to study negative-ion surface production in cesium-free plasmas[30] for many reasons. HOPG (Highly Oriented Pyrolitic Graphite) is chosen as a reference material in our studies as it can be easily cleaved and its yield of NI production is relatively high[31]. Another material of major interest for NI surface production is diamond. Given its negative electron affinity when it is hydrogen terminated and its variable wide energy band gap (depending on the doping), diamond presents electronic properties that may be advantageous for negative-ion production[32]. A significant enhancement of NI yield on boron-doped diamond at high temperature (400-500°C) has been shown earlier[33]. In the present paper a microcrystalline boron-doped diamond (MCBDD) layer of 20 μm thickness produced by plasma-enhanced chemical vapor deposition is used.

The samples are introduced in the RF plasma discharge and negatively biased with respect to the plasma potential[34]. Positive ions are attracted by the surface bias, some of them are backscattered and subsequently converted into negative-ions when leaving the surface. If the surface is hydrogenated, physical sputtering of adsorbed hydrogen may also contribute to the production of negative-ions. NI formed on the surface are accelerated by the sheath in front of the sample and "self-extracted" on the other side of the plasma discharge towards a mass-spectrometer (MS) placed opposite the sample 37 mm away. Negative ions are detected according to their energy and mass, and Negative-Ion Energy Distribution Function (NIEDF) is measured. A dedicated model has been developed in order to take into account transmission of negative ions through sheaths, plasma and mass-spectrometer[34,35,36]. It allows to determine the distribution function in energy and angle of the negative-ions (NIEADF) emitted by the sample from the MS measurements. In previous publications the surface bias of the sample was fixed at -130 V. Consequently, impact energy of the incident particle was 45 eV per proton or deuteron since the ion population was largely dominated by $H_3^+$ or $D_3^+$ ions which dissociate at impact, and because the plasma potential was around 5 V[36]. In the present work we study the influence of the ion energy on the NI surface production down to 10 eV/nucleon. A special focus is put on low bias exposure as it is relevant to low ion/atom impact energies on cesiated grid in real NI sources for NBI systems. We first study the possibility to self-extract the ions at low bias. We then focus on interpretation of negative-ion yield variations with bias. Measurements are conducted at low positive ion flux compared to negative-ion sources (~$5\times10^{13}$ ions cm$^{-2}$s$^{-1}$ versus ~$5\times10^{16}$-$1\times10^{17}$ ions cm$^{-2}$s$^{-1}$ [28,29,37]) in order to limit thermal drifts and erosion issues, as well as to allow for simple time resolved measurements.

The paper is organized in two parts. In the first one, the experimental set-up is briefly described. In the second one the experimental results are first presented with emphasis on the NI yield variations with bias. Second, theoretical considerations, modelling results, as well as complementary experiments are used to interpret NI yield variations with bias for both materials, HOPG and Diamond.

## 2. Experimental set-up

The reactor and diagnostics used are described in detail elsewhere[34-36]. Measurements are performed in a spherical vacuum chamber with a radius of 100 mm. Plasma discharge is created in a Pyrex tube on top of the chamber with RF power (13.56 MHz) applied to an external antenna. A sample is placed in the centre of the spherical chamber thanks to a molybdenum substrate holder. The sample surface exposed to plasma is a disc of 8 mm in diameter facing the mass spectrometer entrance located at 37 mm away.

Microcrystalline boron-doped diamond (MCBDD) layers of 20 µm thickness deposited on doped silicon by plasma-enhanced chemical vapor deposition at LSPM laboratory have been used in the present study. Raman spectroscopy characterization of layers deposited in identical conditions can be found in reference 30. The boron doping level is quite high leading to an electrical conductivity good enough to ensure that the diamond surface bias is identical to the DC bias applied. The HOPG material studied was of ZYB type purchased from MaTeck GmbH company. The density and electrical resistivity of HOPG were 2.265 g·cm−3 and 3.5×10−5 Ω·cm, respectively.

Measurements are performed at 2 Pa $D_2$, 20 W of injected power. The discharge is operated in the capacitive coupling regime. Such low level of power has been chosen to avoid perturbations due to plasma potential fluctuations which affect NIEDF measurements. In order to further reduce perturbations a grounded metal plate has been installed 50 mm above the MS and the sample[34]. Plasma parameters have been measured thanks to a RF compensated Langmuir probe from Scientific System equipped with a 10 mm long 200 µm in diameter tungsten tip, located at about 1.5 cm from the backside of the sample holder. Only estimated values of electron density $n_e$ and electron temperature $T_e$ are given here due to the difficulty to obtain efficient RF compensation and good signal over noise ratio in very low density hydrogen RF plasma[38]. The electron density is estimated to be in the range $10^{13}$-$10^{14}$ m$^{-3}$ and the electron temperature about $T_e = 3.5$ eV, giving an estimated positive-ion flux of 1 to 10 µA/cm$^2$. Due to the importance of knowing the positive ion flux in the present study, a more accurate measurement has been performed. The positive ion flux onto the sample has been measured versus the sample bias by isolating completely the sample from the sample holder. In such way the sample current could be measured independently from the sample holder current, and the sample holder was serving as a guard ring during the measurement. The positive ion current on the sample was found to increase from 2.5 µA (5µA/cm$^2$) at -10 V to 5.7 µA (11.4 µA/cm$^2$) at -170 V.

The positive ion flux repartition determined by mass spectrometry is $D_3^+$ (~80%) followed by $D_2^+$ (~18%) and $D^+$ (2%). The sample can be biased to negative voltages in order to get the positive ion impact energy of ~10−60 eV per deuteron if one restricts the analysis to $D_3^+$ ions. The plasma potential $V_p$ in the vicinity of the sample holder depends on surface bias $V_s$ and ranges from 18 V at $V_s = 0$ to 5 V at $V_s = -170$V. It was determined by using mass spectrometer positive ion energy distributions measured as a function of surface bias. Positive ion distribution peak positions were always easily identifiable (see reference 36 for an example) and were associated to the value of the plasma potential. This measurement was in agreement with the Langmuir probe measurements within few volts. The impact energy of the dominant incident ion $D_3^+$ can be estimated as $e(V_p-V_s)/3$ per deuteron. All measurements have been done with sample surface at room temperature.

## 3. Results and discussion

### 3.1 Negative Ion Energy Distribution: experiments and modelling

A set of measured NI distribution functions for HOPG and MCBDD surfaces exposed to $D_2$ plasma at different surface biases $V_s$ is shown in Figure 1 and Figure 2. A relatively high and noiseless signal can be acquired even with a bias as low as -10 V. In this situation the self-extraction is ensured by the 10 V difference between the sample surface and the mass spectrometer entrance. At $V_s$ below 10 V measurements are difficult as the signal level is strongly decreasing. The maximum value of the NIEDFs and its evolution with the surface bias differ noticeably for HOPG and MCBDD. On the contrary, the shapes of NIEDFs are quite similar (see Figure 3). The NIEDFs include low-energy peak (in the region 1−5 eV) and higher energy part that extends up to an impact energy defined by the positive ion impact energy minus the minimum energy transfer to the solid[31]. It has been shown earlier that the negative ions are formed by backscattering of positive ions as NI and by sputtering of adsorbed hydrogen (deuterium) atoms as NI[4,6]. A model has been developed [34,35,36] to interpret the shape of the NIEDF. This model is presented afterwards.

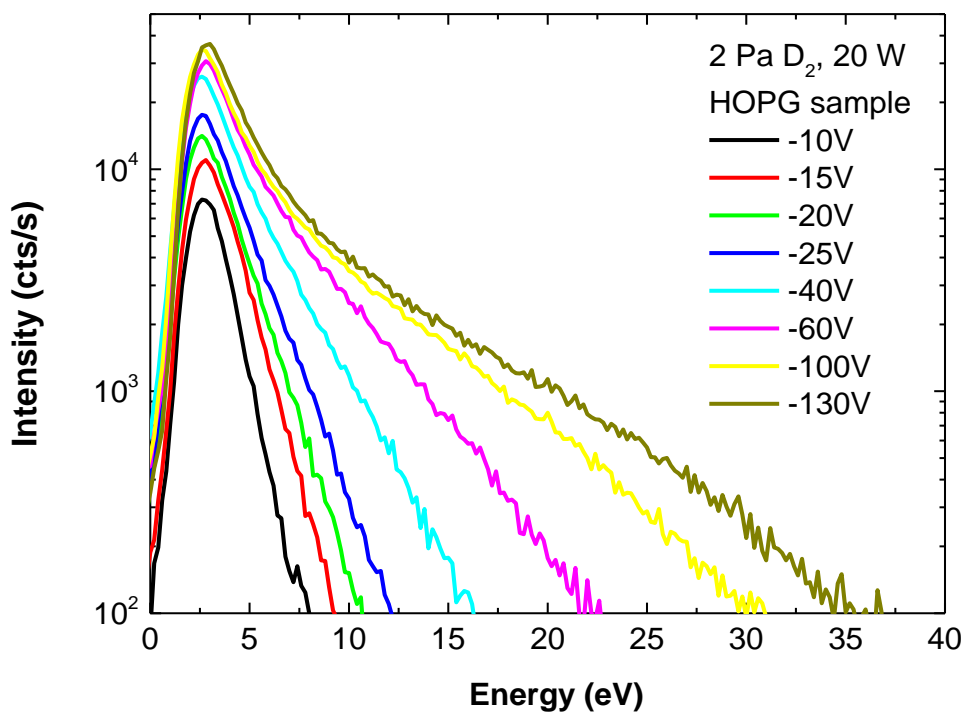

Figure 1. NIEDF measured for HOPG sample exposed to 2 Pa $D_2$ 20 W RF plasma at different surface biases at room temperature.

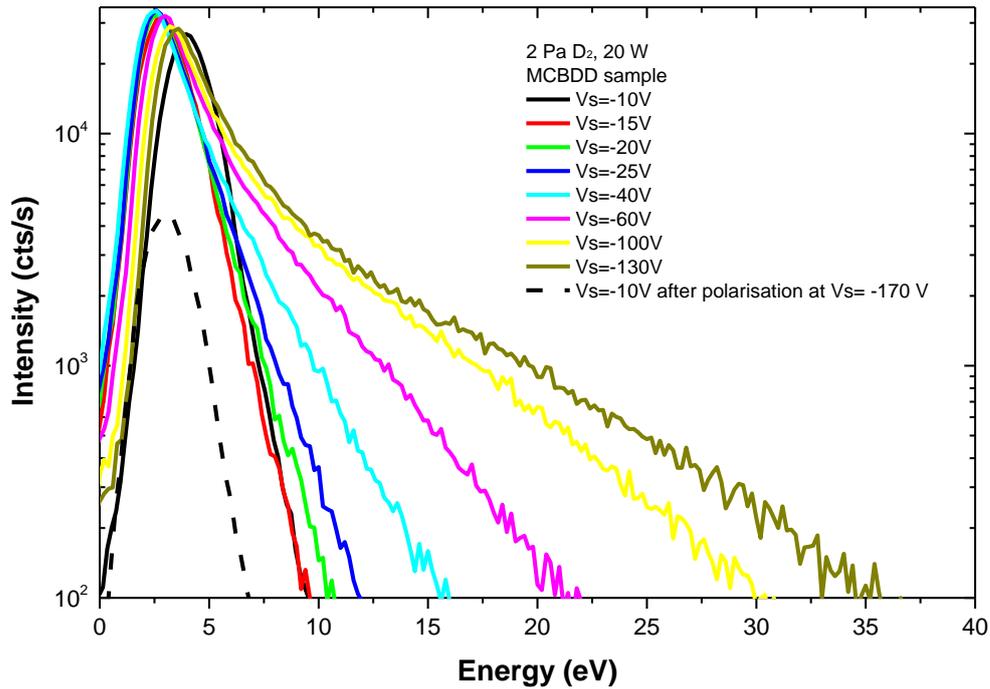

*Figure 2: NIEDF measured for MCBDD sample exposed to 2 Pa D$_2$ 20 W RF plasma at different surface biases at room temperature.*

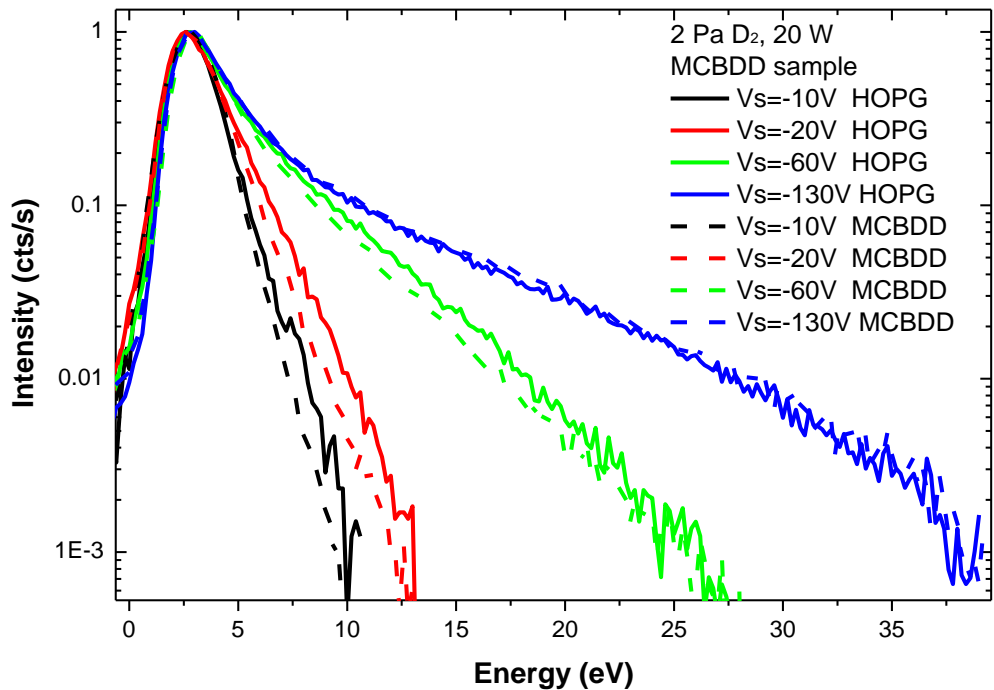

*Figure 3: Normalized NIEDF measured for MCBDD and HOPG sample exposed to 2 Pa D$_2$ 20 W RF plasma at different surface biases at room temperature. MCBDD NIEDFs at -10V and -130V have been slightly shifted on the horizontal axis to match the maximum of all NIEDFs and favour the comparison of the shapes.*

The yield of negative-ions measured for a given angle θ and energy E of emission is:

*Eq. 1*

$$Y_{NI}(E,\theta) \propto P_{iz}(E,\theta,V_s) \times \left(Y_{sp}(E,\theta,V_S) + Y_B(E,\theta,V_S)\right) \times T_{pl}(E,\theta,V_S) \times T_{MS}(E,\theta,V_S)$$

The yield is defined as the flux of negative-ions divided by the total flux of positive ions impinging on the surface. $Y_{sp}(E,\theta,V_S)$ and $Y_B(E,\theta,V_S)$ are the sputtering and backscattering yields for particles leaving the surface with an energy E and angle θ, upon the bombardment of positive ions at energy $E_0 = e(V_P-V_s)$. $P_{iz}(E,\theta)$ is the ionization probability of such particles (the probability of their conversion into negative-ions). It is assumed constant for any E and θ (discussed afterwards) but might still depend on $V_s$ through a change of surface state due to the change of impinging ion energy : $P_{iz}(E,\theta,V_s) = P_{iz}(V_s)$. $T_{pl}(E,\theta,V_S)$ and $T_{MS}(E,\theta,V_S)$ are the transmission probabilities through the plasma and through the mass spectrometer respectively. The first one tells if the emitted ions can reach the mass spectrometer entrance with an angle below the acceptance angle, it is 0 or 1. The NI trajectory depends on the emission energy and angle ($E$ and $\theta$) and on the electric field in the sheath, which is set by the sample bias ($V_s$). The acceptance angle depends on the arrival energy of the negative-ion at the mass spectrometer $E_{MS}$ which is set by both emission energy and sample bias:

$$E_{MS} = E + e(V_{MS} - V_s) = E - eV_s$$

$T_{MS}(E,\theta,V_S)$ is the transmission probability inside the mass spectrometer. It is calculated using SIMION software[39]. It depends on the arrival angle and energy at the mass spectrometer. Most of our recent calculations averages this transmission over the arrival angles from zero to the acceptance angle. Therefore $T_{MS}(E,\theta,V_S) \approx T_{MS}(E,V_S) = T_{MS}(E_{MS})$. For more details on the transmission functions, see references 34 and 36.

In the present experiments the sample bias is varying between -10 V and – 170 V giving an impact energy per deuteron ($D_3^+$ is the dominant positive ion in the plasma) between ~10 eV and 60 eV taking into account the plasma potential and its changes versus sample bias. The mass spectrometer measurement gives a negative ion distribution function f(E). The model computes this distribution function by assuming the ionization probability constant ($P_{iz}(E,\theta) = 1$) and integrating Eq. 1 over all the angles θ:

*Eq. 2*

$$f''(E) \propto \int \left(Y_{sp}(E,\theta,V_S) + Y_B(E,\theta,V_S)\right) \times T_{pl}(E,\theta,V_S) \times T_{MS}(E,V_S) \, d\theta$$

The NIEDF calculated by the model without taking into account $T_{MS}(E,\theta,V_S)$ is labelled f':

*Eq. 3*

$$f'(E) \propto \int \left(Y_{sp}(E,\theta,V_S) + Y_B(E,\theta,V_S)\right) \times T_{pl}(E,\theta,V_S) \, d\theta$$

f'' is more accurate than f' but requires time consuming calculations to get $T_{MS}$. We often use f' to simplify the analysis. We have noted previously that the differences between f' and f'' are not huge, and f' calculation is often enough to interpret the shape of NIEDF. The transmission through the MS mostly modifies the signal intensity. In the equations above $T_{pl}(E,\theta,V_S)$ is calculated from Newton's law of motion based on the known electric field in the sheaths and on NI emission energies E and angles θ. The experimental arrangement ensures planar sheaths in front of sample and mass spectrometer. Therefore, the electric field can be obtained from the Child Langmuir law knowing plasma parameters thanks to Langmuir probe measurements. The difficulty of the model is to estimate backscattering and sputtering yields, $Y_{sp}(E, \theta, V_S)$ and $Y_B(E, \theta, V_S)$, as a function of energy E and angle of emission θ. SRIM software [40] has been used to compute the energy and angle distributions of backscattered and sputtered particles $Y_{sp}(E, \theta, V_S)$ and $Y_B(E, \theta, V_S)$. SRIM is a Monte-Carlo code based on the binary collision approximation (BCA) which assumes that collisions between atoms can be approximated by binary elastic collisions described by an interaction potential. As discussed by Eckstein et al.[41] the validity of the assumptions behind the BCA is expected to gradually decrease below ~30 eV. Our model has been mostly used up to now to analyse data obtained at $V_s$ = -130 V[34,35,36]. Under this condition the positive ion energy is around 50 eV/nucleon and SRIM assumptions are probably fulfilled giving a very good agreement between calculation and experiments[30,36] when using the SRIM input parameters listed in reference 42. In the present paper lower positive ion energy (lower bias in absolute values) are explored and SRIM cannot be used, or at least SRIM results must be taken with care.

Results of the model are given in Figure 4 showing the modelled NIEDFs at the mass spectrometer (f' from Eq. 3) and the NIEDFs on the surface $f_{mod}$ as computed by SRIM:

*Eq. 4*

$$f(E) \propto \int \left( Y_{sp}(E, \theta, V_S) + Y_B(E, \theta, V_S) \right) d\theta$$

An important outcome of the model is that the fraction of emitted NI which is collected by MS is relatively small. Many ions miss the entrance of the mass spectrometer or reach it with an angle $\theta_{MS}$ higher than the acceptance angle $\theta_{aa}$. In particular ions emitted at high energy are not efficiently collected since their trajectories are not sufficiently rectified by the electric field in the sheath and they arrive at mass spectrometer with a too high angle[35].

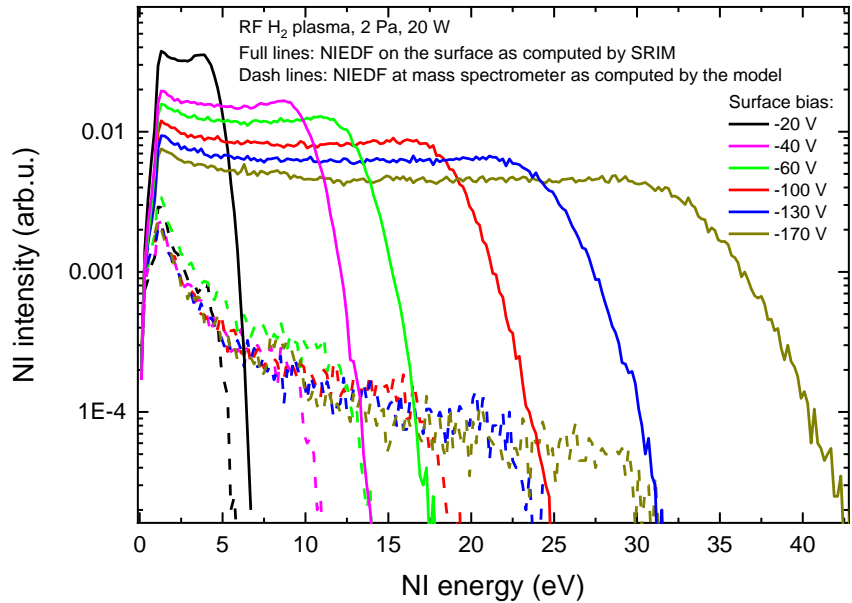

*Figure 4: NIEDFs on the sample surface (full lines) computed by SRIM. NIEDF at the mass spectrometer (dash lines) computed using SRIM results and NI trajectory calculations. Parameters for the calculations are those of a $D_2$ RF plasma 20 W, 2 Pa.*

All SRIM calculations have been made assuming a 30% constant deuterium percentage on the surface whatever the bias. This choice, validated at high bias ($V_s = -130$ V), is arbitrary here. Two sets of calculations have been performed, one with a layer density of 2.2 g/cm$^3$ and one with 3.5 g/cm$^3$ corresponding to deuterated HOPG and deuterated diamond respectively. No major difference was observed between both sets of calculations, neither on the yields nor on the NIEDF shapes. Therefore, only one set of calculation is presented (2.2 g/cm$^3$) in Figure 4. Comparison of model and experiment with HOPG material is given in Figure 5. Distribution functions are normalized and only shapes can be compared in this figure. Obviously, the differences between f' and f'' are light and f' can be used to study the shapes of NIEDF. The agreement between experiments and calculations at $V_s = -130$ V, if not perfect, is quite satisfactory. One must keep in mind that in RF plasma the NIEDF are slightly broadened by the RF fluctuations of the plasma potential and therefore the model never perfectly matches to the experiment[34]. However, the validity of the model has been proven using microwave plasma (ECR excitation) [34]. It is also possible to improve the agreement between experiment and modelling by taking into account the relative ratio of $D_3^+$, $D_2^+$ and $D^+$ ions as well as their real energy distribution as demonstrated in a previous paper[36]. However, this is not the goal of the present paper and to speed up the calculations only mono-energetic $D_3^+$ ions have been considered for the modelling.

It has been shown that changing $Y_{sp}(E, \theta, V_S)$ and $Y_B(E, \theta, V_S)$ or changing hydrogen coverage noticeably affects the computed distribution[34,35] at $V_s = -130$ V. However, as it can be seen from Figure 5, for the low bias case, changing the hydrogen coverage does not affect strongly the computed NIEDF. The agreement between model and experiment at low bias is therefore only showing that angle and energy distribution functions computed by SRIM are roughly correct. There is not enough sensitivity of the model to the input parameters to fully validate SRIM calculations at low positive ion energy (low bias in absolute values). One more difficulty in analysing the low bias results arises when considering the total NI counts (integral of NIEDF) rather than shapes of NIEDF. This is detailed in next paragraph.

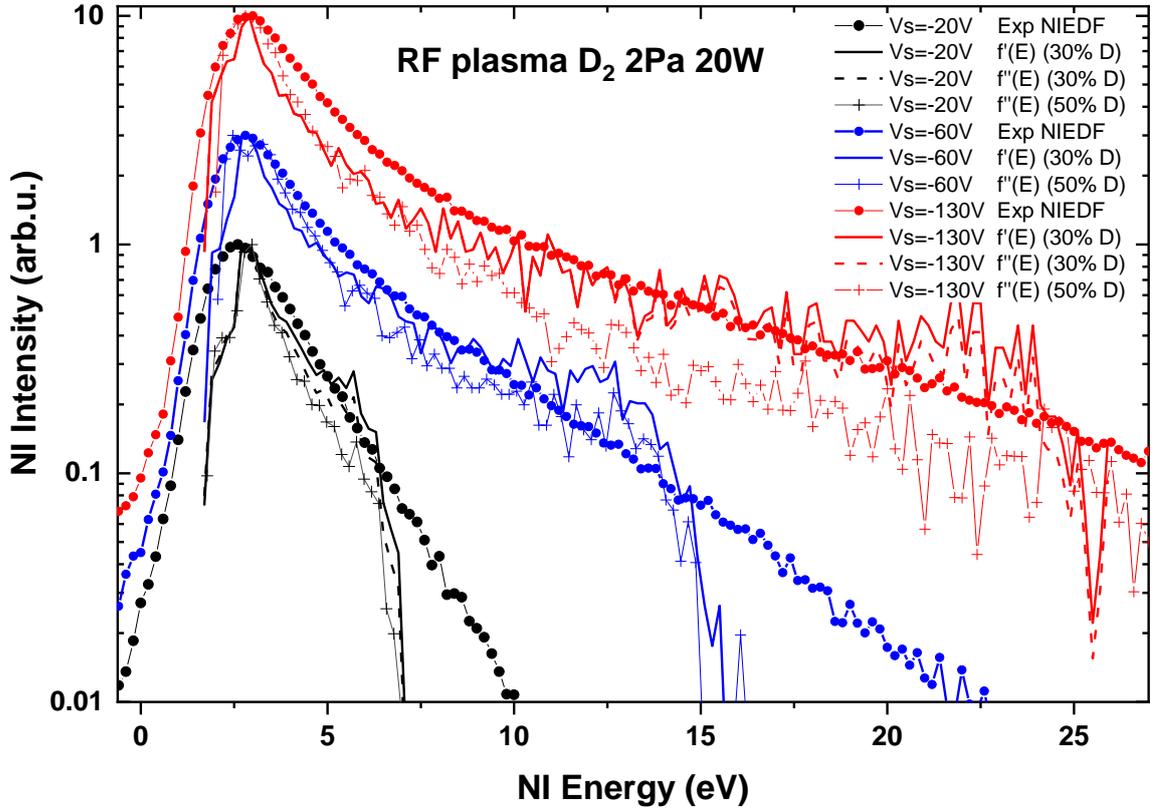

*Figure 5: comparison between experimentally measured NIEDFs (line plus symbols) with HOPG sample and computed ones (lines). f''(E) and f'(E) stand for NIEDFs computed with or without the mass spectrometer transmission function. Experimental conditions are $D_2$ RF plasma 20 W, 2 Pa, bias surface of -20 V (results normalized to one), -60 V (results normalized to three) and -130 V (results normalized to ten for sake of clarity). Considering input parameters of calculations, the deuterium content on the surface is indicated in the figure and the impact energies used are 10eV for Vs=-20V, 25eV for Vs=-60V and 45eV for Vs -130V. At -60V normalized functions f' and f'' cannot be distinguished since TMS ≈ 0.8 = constant for any ion energy below 15 eV.*

### 3.2 Negative Ion yields

Figure 6 shows measured NI yields as a function of $V_s$ for MCBDD (filled red squares) and HOPG (filled black circles) in $D_2$ plasma. The measured yield is defined as the ratio between the measured NI total flux (integral of the measured NIEDF) and the positive ion flux. It is given in arbitrary units since the mass spectrometer signal scale is not absolutely calibrated. It can be seen that both materials demonstrate different behaviour. The yield is increasing for HOPG while it decreases for MCBDD when $V_s$ is going from -10 V to -170 V. MCBDD is the best NI producer at biases between -10 V and -60 V (impact energy below 25 eV). The open symbols in Figure 6 show NIEDF peak intensities normalized by the PI current as a function of surface bias. The NIEDF peak intensity is used as a representation of the yield of low energy emitted NI (0 – 5 eV) since the peak is always located between 0 and 5 eV whatever the surface bias. Peak variations allow to infer information on surface ionization as it will be shown in paragraph 3.3. The peak intensity is stronger for MCBDD between -10 V and -60 V and higher for HOPG between -60 V and -170 V, and the signal is decreasing with $V_s$ for MCBDD, while it is increasing with $V_s$ for HOPG. The measurements were done on a pristine sample starting at $V_s$ = -10 V and proceeding till -170 V. Let us note that all measurements were done at steady state when the yield stabilized. It required usually few minutes at each bias[32].

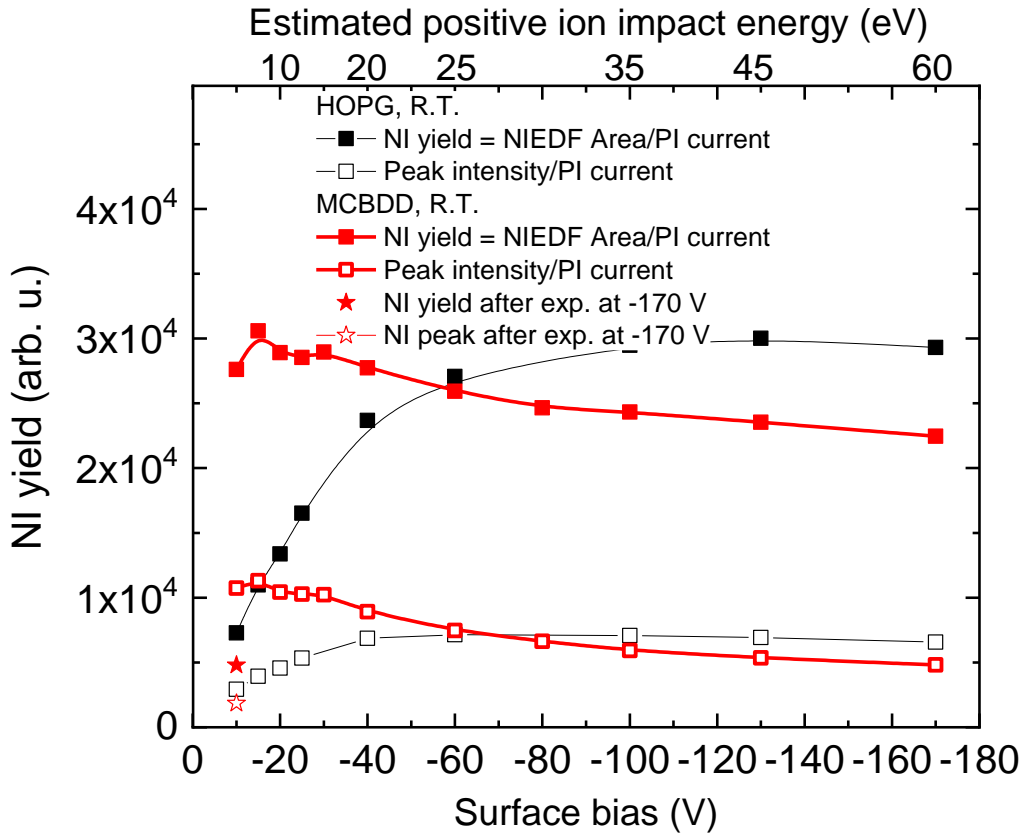

Figure 6. NI yield and peak intensity of NIEDF measured for MCBDD and HOPG exposed to 2 Pa $D_2$ 20 W RF plasma at room temperature versus surface bias. The estimated positive ion impact energy for each bias is indicated on top. The relation between impact energy and bias is not direct due to the plasma potential changes with surface bias.

The use of SRIM assuming an amorphous a-C:D (30%) top layer for both diamond and graphite is immediately questioned by the results presented in Figure 6 since both materials behave completely differently in terms of yields. The variations of yield computed by SRIM with a-C:D (30%) top layer are presented in Figure 7. If one neglects the transmission function of the mass spectrometer $T_{MS}$ the yield slightly increases with bias but not as much as observed for graphite in Figure 6, and the peak intensity slightly decreases contrary to graphite in Figure 6. When considering $T_{MS}$, yield and peak intensity globally decrease with bias which is in contradiction with graphite behaviour in Figure 6.

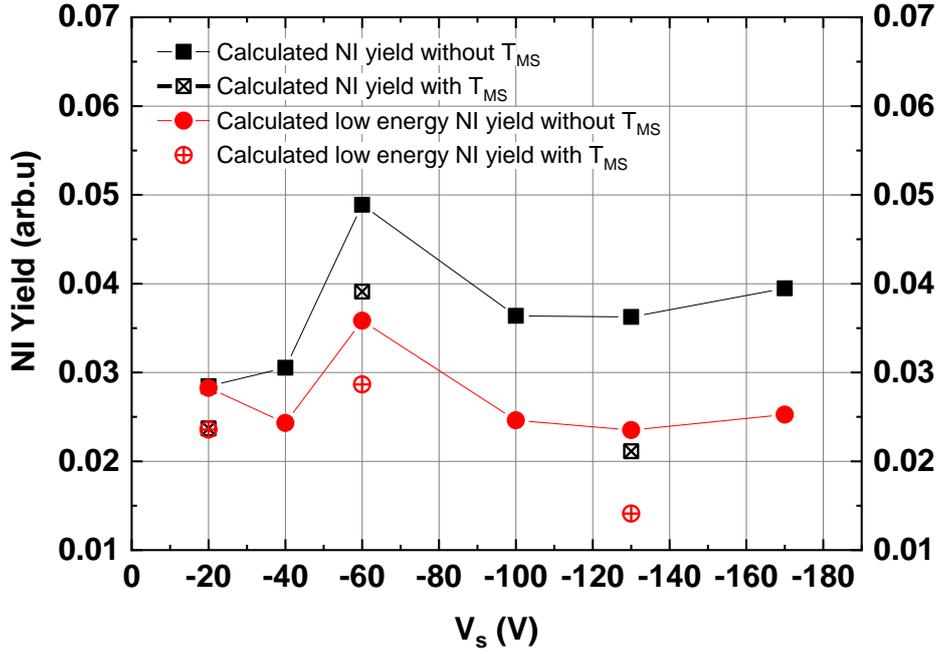

*Figure 7: Calculated NI (square) and low energy NI (circle) yields. Filled/open symbol are for calculations without/with taking into account $T_{MS}$. Parameters for the calculations are those of a $D_2$ RF plasma 20 W, 2 Pa.*

In Figure 3 it is shown that NIEDFs on HOPG and MCBDD have similar shapes. It suggests that for both materials the normalized angle and energy distributions of emitted negative ions (NIEADF) are globally similar. As it has been shown that NIEDF shapes obtained at $V_s$ = -130V are determined by the relative contribution of backscattering and sputtering processes to the negative-ion emission, it also suggests that this relative contribution is identical for both materials, at least at -130V. Indeed, materials with different backscattering and sputtering contribution lead to marked differences in NIEDF shapes as demonstrated in 36.

### 3.3 Interpretation of negative ion yield variations with bias

In order to interpret yields behaviour with $V_s$ variations, let us come back to Eq. 2 dropping any assumption on the ionization probability:

*Eq. 5*

$$f''(E) \propto \int P_{iz}(E, \theta, V_s) \times \left(Y_{sp}(E, \theta, V_S) + Y_B(E, \theta, V_S)\right) \times T_{pl}(E, \theta, V_S) \times T_{MS}(E, V_S) \, d\theta$$

If one assumes that NIEADF are globally similar for both materials, then the plasma and mass spectrometer transmission functions $T_{pl}(E, \theta, V_S)$ and $T_{MS}(E, V_S)$ are identical for both materials. Nevertheless, it is instructing to study their variations with surface bias. To do so, the fraction of collected ions has been computed versus the surface bias using the model with a uniform negative-ion energy and angle distribution function (NIEADF) on the surface. The uniform distribution is simply defined by $f(E, \theta) = 1 \; \forall E, \forall \theta$. The fraction of collected ions is the total number of collected ions (integral of f'(E) or f''(E) distribution between 0 and $E_0/3$) divided by the total number of emitted ions (integral of f(E) distribution between 0 and $E_0/3$).

The fraction of low energy collected ions (represented in the experiment by the peak intensity of the NIEDF) has been also computed by integrating f, f' and f'' between 0 and 5 eV. The results are presented on Figure 8. First of all, it has been checked that all negative ions collected emerged from the sample surface and no negative-ion can be collected originating from the surrounding surfaces such as the clamp. Indeed, the model shows that all collected negative-ions originate from a spot on the sample surface (~2 mm in diameter) which does not exceed the sample dimensions (8 mm in diameter) for any $V_s$ studied here, even at low bias. The calculations also show that only about 4-8 % of the emitted ions are measured, this fraction being higher if one considers the low energy ions rather than the whole NIEDF (12-15 %). The better collection efficiency of low energy ions was already observed and explained[34].

Without taking into account the transmission function of the mass spectrometer $T_{MS}$, the extraction efficiency is almost constant with surface bias. When increasing the bias in absolute values, the arrival energy of the negative-ion at the mass spectrometer $E_{MS}$ is increasing leading to a decrease of the acceptance angle. However, at the same time the higher electric field in the sheath tends to decrease the NI arrival angle at the mass spectrometer $\theta_{MS}$. Both effects compensate to give almost constant extraction efficiency when neglecting $T_{MS}$. However, when $E_{MS}$ is increasing $T_{MS}$ is decreasing due to the difficulty to focus high energy ions inside the mass spectrometer. Therefore, the extraction efficiency is globally decreasing when increasing the bias. It can be noted that it is worth taking into account $T_{MS}$ to analyse yield variations while NIEDF shapes can be analysed ignoring $T_{MS}$ (Figure 5).

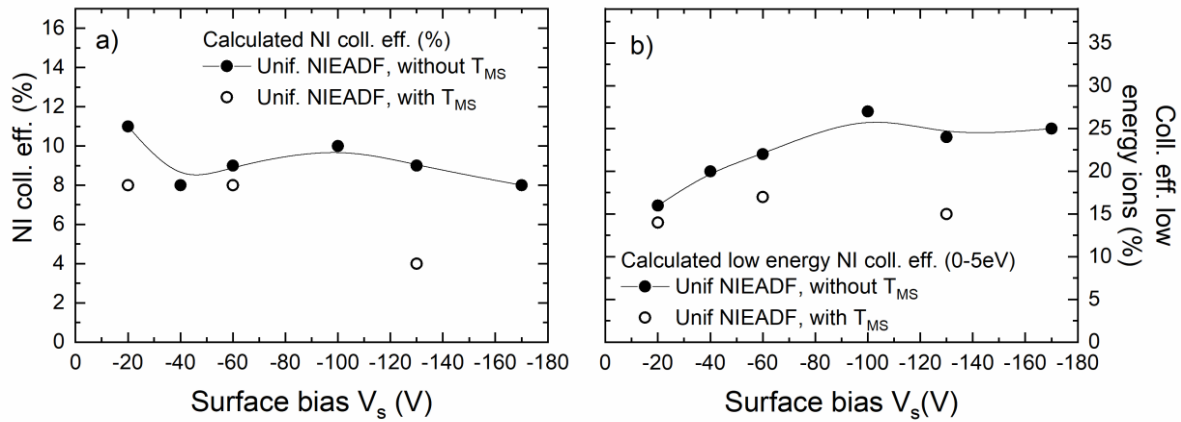

Figure 8: Collection efficiency computed by the model using a uniform NIEADF. The filled/open symbols are for calculations without/with the transmission function of the mass spectrometer taken into account. Parameters for the calculations are those of a $D_2$ RF plasma 20 W, 2 Pa. a) results for the full energy range of NI b) results limited to NI emitted with energy between 0 and 5 eV

The collection efficiency for low-energy ions (0 – 5 eV) is increasing when ignoring $T_{MS}$. The decrease of the acceptance angle is compensated by the fact that the trajectories of low energy ions are easily modified by the electric field in the sheath leading to $\theta_{MS}$ well below the acceptance angle. This explains their better collection efficiency. If one considers now the calculations taking into account the mass spectrometer transmission function, the collection efficiency is more or less constant for low energy NI. The conclusion to this part is that the collection efficiency is expected to be at best constant or to decrease over the whole bias range. Any increase of the yield such as observed for HOPG material cannot be interpreted by an increase of the NI extraction and collection efficiency.

It is not easy to get direct information on the ionization probability from plasma experiments such as those presented here. For metals, the ionization probability is expected to depend on the perpendicular velocity of the outgoing particle. Here, when the bias is increasing in absolute values, the mean energy of the D particles leaving the surface is also increasing as shown by the NIEDF tail which is expending to higher energy in Figure 1 and Figure 2. Such increase of the mean energy could lead to an increase of the ionization probability. In order to rule out this possibility we have plotted the peak intensity variations in Figure 6. NIEDF peak is always located around 3-4 eV and therefore corresponds to NI emitted at constant low energy whatever the surface bias. For such ions, the ionization probability cannot increase because of an increase of the perpendicular velocity. As the peak intensity variation is following the yield variation, we assume that the increase of the mean NI energy is not playing a crucial role and is not affecting strongly the ionization probability. We can maintain the assumption we made in previous works of no dependence with energy and angle of the ionization probability for graphite and diamond[32,34,35,36].

The ionization probability could still change with sample bias if the surface state is changing. Indeed, the ionization probability is defined for one material with given electronic properties. Changing the electronic properties changes the ionization probability. In order to investigate this possibility, we have measured NI yields at low bias after high bias experiment. The idea is to measure NI under low bias condition with a surface state corresponding to the high bias condition. If after a complete scan of $V_s$ HOPG is exposed again at low bias, the same NI yield as before is immediately obtained, and no time evolution is observed. The experiments were done with a time resolution on the order of one second which is usually enough to observe surface modifications due to the ion bombardment[32] since the ion flux is rather low in the present experiments. A fluence of roughly one mono layer of material ($10^{15}$ ions cm$^{-2}$) is reached after about 20 seconds. Let us note that a complementary experiment has been performed in which after high bias exposure, the bias has been modulated between -100 V (duration 5 ms) and -20 V (duration 50 µs). In this case the sample is exposed to high bias most of the time. The mass spectrometer acquisition was performed during the low bias (-20 V) phase. The yield at -20 V was the same during the modulated bias experiment as the initial yield at -20 V before high bias exposure. It clearly shows that when switching from high bias to low bias, the yield comes back immediately to its initial low bias value without any time evolution. As the surface state, determined by parameters such as the sp$^2$ over sp$^3$ ratio or the deuterium content, cannot change fast, it demonstrates that the change of surface state between low and high bias, if any, is not promoting the ionization probability for HOPG material. For HOPG, the ionization probability can be considered as a constant over the whole bias range.

In the case of MCBDD its surface produces 7 times less NI at $V_s$ = -10 V after exposure to -170 V (see Figure 6). The initial signal is recovered after several tens of minutes of exposure at -10 V. Same results are obtained with – 20 V bias. This is a remarkable difference between both materials. When the bias is increased in absolute values the energy of the positive ions impacting on the surface and creating defects is increased, leading to i) a deeper ion implantation and a thicker defective layer on top of the pristine material ii) possibly creation of defects of different nature in the defective layer. These effects can be summarized as a change of the surface state. This change is obviously unfavourable for NI surface production on MCBDD since after a complete scan in bias, the signal is strongly decreased when coming back to -10 V. The ionization probability of the surface exposed at high bias is obviously lower than the ionization probability of the diamond surface exposed at low bias. This conclusion is in line with previous papers showing that defect creation on diamond is unfavourable for NI production[43,32]. Electronic properties of diamond are promoting NI surface production, defect creation is modifying electronic properties and lowering NI yield.

Unfortunately, there is no in-situ surface analysis in the present set-up to track surface state changes. Some measurements and MD simulations from the literature can be used to interpret experimental results despite they do not match exactly with the present experimental conditions. Davydova et al[44] have studied the impact of $H^+$ ions on multi-layer graphene at increasing energy. At energy below the penetration threshold through the first basal plane (5 eV) ions hydrogenate the top surface. At 10 eV $H^+$ ions have enough energy to penetrate through the first plane but not through the second. They create a strongly disordered and hydrogenated top layer and etching starts. Once the a-C:H layer is formed $H^+$ ions lose less energy to pass through it and start to hydrogenate the third and fourth layers. An equilibrium is reached between erosion and hydrogenation. At 25 eV the process is globally identical, but the first three layers are initially hydrogenated by the impact of ions. Bombardment by 25 eV and 50 eV $H_2^+$ ions show similar trend to 10 eV and 25 eV $H^+$ bombardment since a large fraction of $H_2^+$ ions dissociate at impact and the energy is shared between the fragments. From Davydova et al work it is seen that the number of CH, $CH_2$ and $CH_3$ bond ratio in the defective layer is changing with the positive ion energy, as well as the hydrogen content which is increasing with the ion energy. However, this defective layer seems to be highly hydrogenated and porous with electronic properties similar to those of a soft a-C:H whatever the positive ion energy is. It might explain why the ionization probability is found constant. The consequence of an increase of positive ion energy would be to reach deeper graphene layers rather than drastically change the top surface layer composition. This would make a huge difference with diamond for which increasing ion energy would help moving carbon atoms from their lattice sites. Dunn et al[45] have compared diamond and graphite irradiation by 15 eV tritium positive ions with MD simulations. They have shown that the rigid structure of diamond maintains carbon atoms in place and the number of atoms in each layer does not change much, despite carbon atoms may lose their $sp^3$ hybridization due to the hydrogenation[45,46]. The parallel layers and less dense structure of graphite allows more mechanical deformation and expansion, which leads to a higher penetration of ions and higher retention. As shown in Dunn thesis[47] the increase of impact energy helps destroying the diamond structure and create a porous and hydrogenated a-C:H film.

Kogut et al[48] demonstrated with MD simulation that the chemical sputtering threshold for diamond at room temperature is around 4 eV. Still with MD, de Rooij found chemical sputtering threshold at 5 eV and penetration threshold for H in diamond around 7 eV[49]. De Rooij also showed that even at 20 eV impact energy H ions went through only one or two carbon layers at maximum. This result is in line with the assumption that impact energy has to be increased well above 20 eV to destroy the diamond structure. However, Yamazaki et al[50] demonstrated experimentally that impact of $H_2^+$ ions on diamond with energy between roughly 10 and 50 eV seems to create a defective layer with constant dangling bond density and CH bonding configuration, but with increasing depth. This observation seems to be in contradiction with the assumption that the increase of impact energy would help destroying the diamond structure and thus changing the CH bonding configuration. It shows that it is hard to infer diamond transformation in the present experimental condition. However we can expect the ionization probability to be affected due to the fact that the top layer composition is changing strongly from diamond like to a-C:D. Let us note that none of the MD simulations referenced above consider the concomitant impact of low energy neutral H atoms with ions while in the experiment the atomic flux is much higher that the ion flux. In reference 51 a MD study of H impact on Si and SiN layer has been conducted with increasing atom to ion ratio. From these results we can expect that H atom impacts on diamond or graphite will increase the top layer hydrogenation and its etching, thus limiting its depth.

After examining all parameters in Eq. 5 we come to the conclusion that HOPG yield increase with surface bias can only be explained by an increase of backscattering and/or sputtering yields $Y_{sp}(E, \theta, V_S)$ and $Y_B(E, \theta, V_S)$. NIEDF in Figure 5 are normalized and the comparison between experiment and calculation gives information on E and θ dependence of yields but not on their absolute value. In SRIM software the surface state is an input of the calculation and the choice of surface state parameters (deuterium content, surface binding energy…) strongly influences the results given. The set of parameters chosen for calculations at $V_s$ = - 130 V, validated in our previous works[36], has been used here for lower bias when its validity cannot be justified. Calculations with this set of parameters show a sputtering yield increasing from 0 at 10 eV impact energy to 0.6% at 50 eV. The backscattering yield is increasing from 13.5% at 10 eV to 15.5% at 25 eV and is then constant. Let us note that despite sputtered particles are in a minority they have a strong influence on measured NIEDF. Indeed, they are emitted at lower energy[42] and are more efficiently collected (Figure 8) than high energy ions. From the model at – 130 V, they represent 30% of the collected NI. Therefore, both sputtering and backscattering yield computed by the model with this set of parameters are increasing with positive ion energy (ie with surface bias in absolute value). This increase is not enough to compensate the decrease of the collection efficiency (Figure 8) and the model is not predicting an increase of NI signal for HOPG. However, let us remember that SRIM calculations at low positive ion energy have to be considered with care and we think that the NI yield increase observed for HOPG can only be explained by an increase of backscattering and sputtering yields with positive ion energy.

This conclusion is reinforced by a complementary experiment. A bias scan using HOPG material was performed in a high density plasma, using the inductive mode (ICP) rather than the capacitive (CCP) power coupling mode of the plasma source. The injected power was 200 W. The measured plasma density was $3.10^9$ cm$^{-3}$, and the positive ion flux was around 100 µA/cm$^2$ (~$6.10^{14}$ ions/cm$^2$s, around ten times higher than for the low density case). Under such high density condition, the sheath in front of the sample is much thinner than in low density conditions and it is not collisional anymore for positive ions. This has been demonstrated in reference 36 where distribution functions of positive ions crossing a high voltage sheath were measured. In high density mode the distribution functions are much more peaked and closer to mono-energetic distributions than in low density conditions. Under such a situation, the NI signal on HOPG is of course higher due to the much higher positive ion flux. But more interesting is the observation of an abrupt increase of NI yield between $V_s = -20$ V and $V_s = -40$ V (see Figure 9). Such threshold effect is most probably corresponding to the onset of physical sputtering. This onset is not seen in low density mode since positive ion distribution functions are more distributed in energy and at any bias a large distribution of positive ion energy is impacting the surface, smoothing the threshold effect. The sputtering threshold explains that the NI yield immediately comes back to its initial value when switching the bias from high value to low value since positive ion energy is immediately lowered.

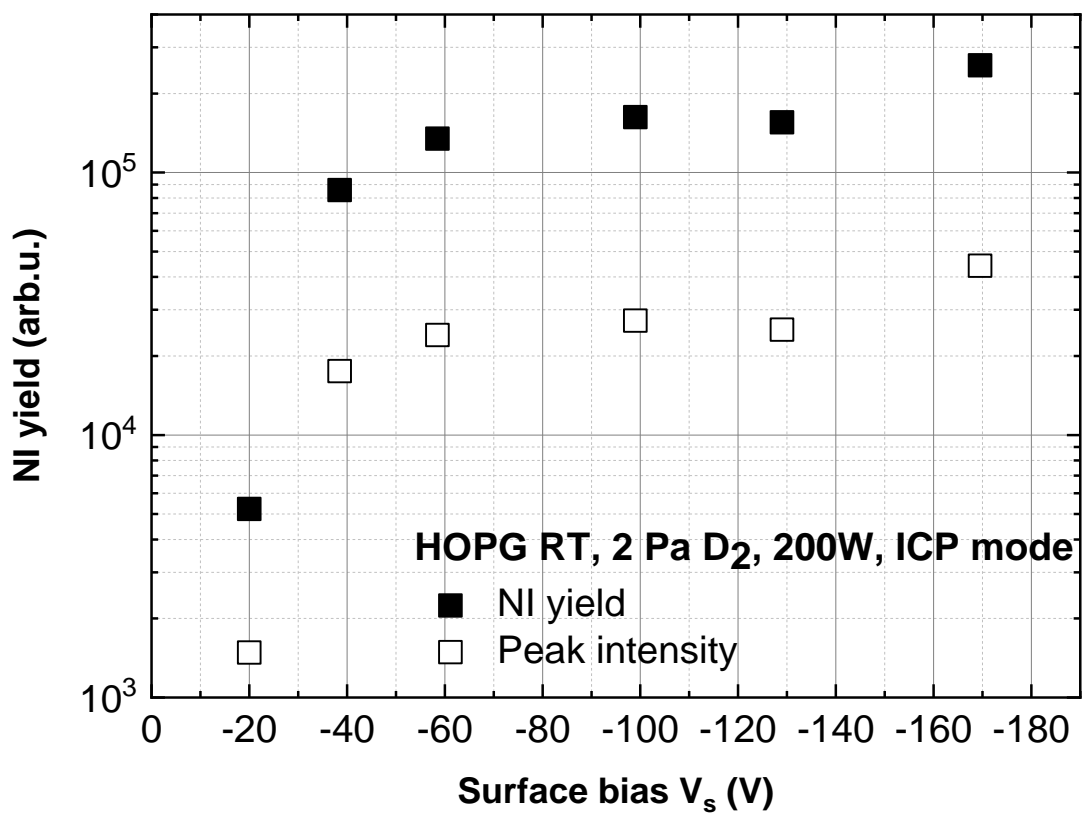

*Figure 9: NI yield and peak intensity of NIEDF measured for HOPG exposed to 2 Pa $D_2$ 200 W ICP plasma at room temperature.*

We therefore conclude that NI yield on HOPG is increasing with surface bias due to an increase of backscattering and/or sputtering yields and despite a decrease of the collection efficiency. This increase of backscattering and sputtering yields cannot be predicted due to the lack of knowledge on the surface state versus the sample bias, and due to the fact that the assumptions behind the SRIM calculations are not fully fulfilled under our experimental conditions. Molecular dynamics simulations could help solving this issue and will be the subject of future works. The same increase of backscattering and/or sputtering yield probably occurs also for diamond but is mitigated by the strong decrease of the ionization probability because of defect creation. For HOPG it has been demonstrated that the ionization probability is not changing with surface bias.

### 3.4 Time variation of negative-ion yields

A way to observe the surface state change of diamond is to take a pristine sample and to trace NI yield in time upon plasma exposure at a given bias. In reference 32 the decrease of NI yield with time at surface bias equal to -130 V was demonstrated and attributed to defect creation and loss of diamond electronic properties. It is shown here that the same process already occurs at low bias (see Figure 10 for MCBDD in a $D_2$ low density plasma with a surface bias of –20 V). It takes about twenty minutes at -20 V to reach steady state. Let us remember that all measurements shown before in the present paper have been obtained at stead state after the initial time evolution of the negative-ion yield. One measurement has been obtained with a sample that was outgassed under vacuum at 400°C prior to experiments. Unfortunately, the first two minutes of time evolution have been lost at acquisition. The initial signal was apparently higher when the sample was first outgassed by roughly a factor 1.2 (extrapolation of the time evolution to time zero). However, no definitive conclusion can be drawn since we used three different samples in three different experiments

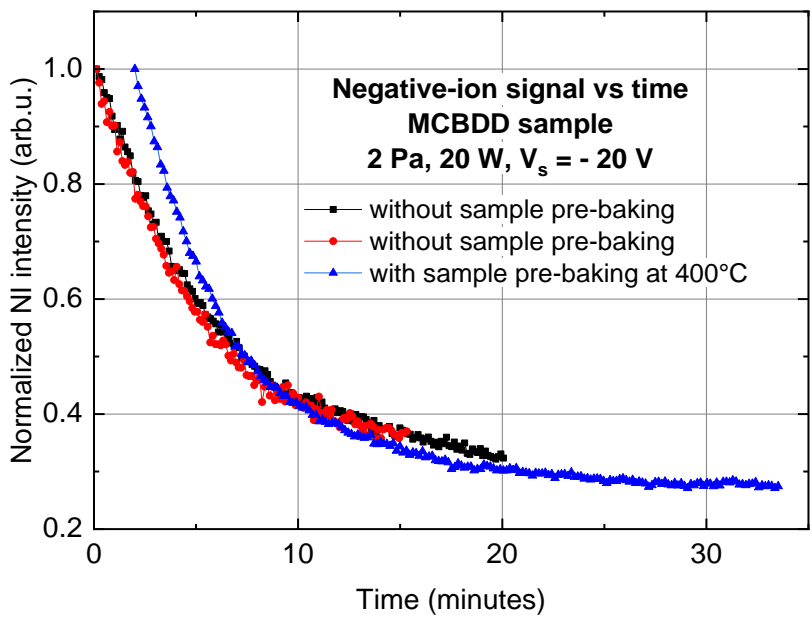

*Figure 10. Time evolution of NI yield produced on MCBDD surfaces: 2 Pa $D_2$ 20W RF plasma, surface bias -20 V. Three different samples in three separate experiments have been used. Results represented by triangle symbols have been obtained with a sample heated under vacuum at 400°C before experiment to desorb any impurity.*

Figure 11 and Figure 12 show NI total signal and NI peak signal variations with time for short exposure duration at each bias for HOPG and MCBDD respectively. In this experiment, the plasma was switched on at -10 V bias using pristine samples. After 30 seconds (20 seconds for MCBDD) of exposure the bias was increased to -15 V and measurements were recorded again for 30 seconds (20 seconds for MCBDD). This procedure was repeated for each bias. The total NI signal is shown by black filled symbols and the NI peak signal is shown by red filled circle symbols. The figures also indicate yield and peak intensity values obtained at steady state in separate experiment.

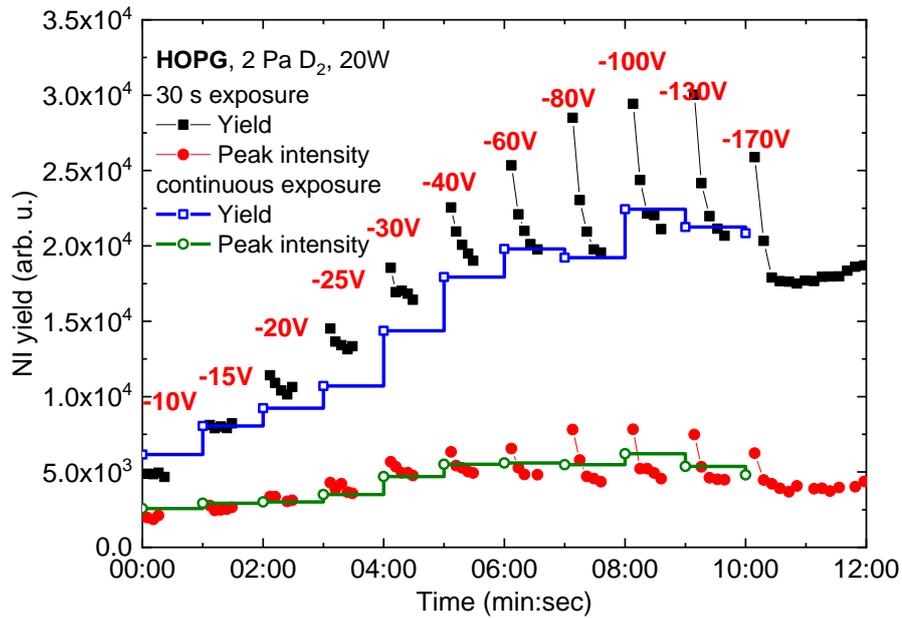

*Figure 11. Time evolution of NI yield and NIEDF peak intensity for HOPG in 2 Pa D$_2$ 20W RF plasma at R.T.: filled symbols refer to a short exposure of 30 s at each surface bias, while open symbols show equilibrium values after long-term continuous exposure.*

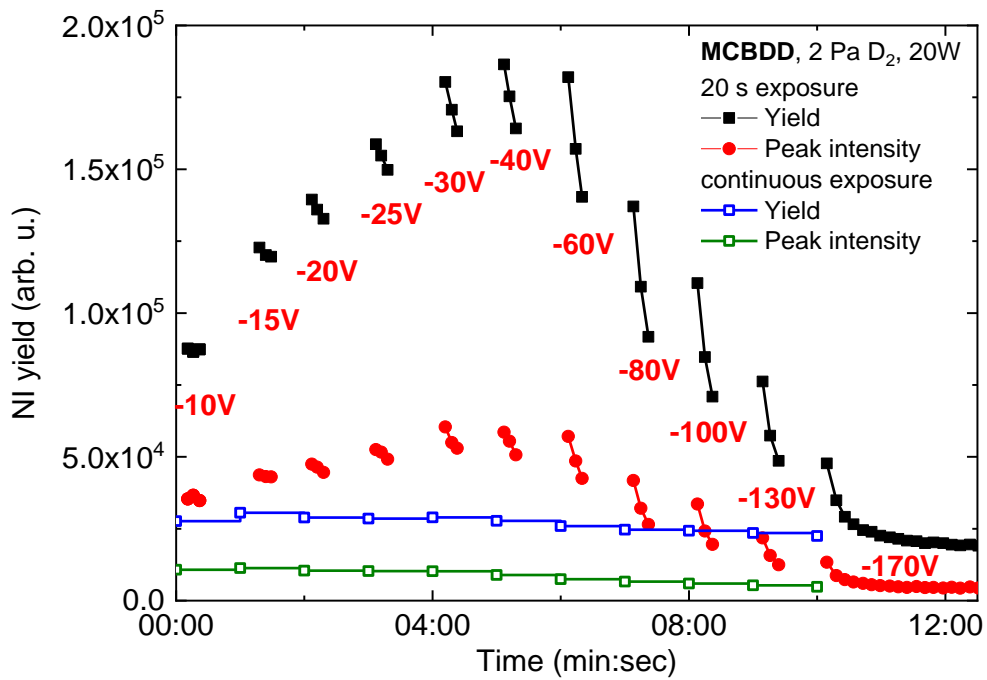

*Figure 12. Time evolution of NI yield and NIEDF peak intensity for MCBDD in 2 Pa D$_2$ 20W RF plasma at R.T.: filled symbols refer to a short exposure of 30 s at each surface bias, while open symbols show equilibrium values after long-term continuous exposure.*

There is one common trend to both materials. When the bias is increased, the signal is initially higher than after 20 or 30 seconds of exposure. So, the increase of positive ion energy leads to a temporary increase of the signal. We attribute this effect to a temporary increase of the backscattering yield due to the fact that the positive ions reach a depth where no previous hydrogen implantation has occurred. In this region the material is the pristine one, made of carbon only, and is denser than the modified one. MD simulations show indeed a decrease of carbon material densities with hydrogen bombardment[52]. SRIM is predicting a decrease of backscattering yield by about 25% when going from a pure carbon layer to a deuterated carbon layer at 50 eV D impact energy. As the material is denser and as the energy transfer from deuterium to carbon is much less efficient that deuterium to deuterium, we can expect that impacting particles on pure carbon have higher chance to reverse their momentum. It explains the higher initial backscattering yield. Then the hydrogen implantation mitigates this effect and the signal comes back to its stationary value for HOPG. Concerning MCBDD the signal is also decreasing with time during the short exposure duration but is still much higher than at steady state. Indeed, the signal at -60 V for instance is initially 7 times higher and still 5 times higher after 20 seconds of exposure compared to its steady state value. It can be observed that the signal starts to decrease at bias higher than -60 V. At -170 V the signal reached its steady state value within the course of the short 20 seconds exposure duration. This is a consequence of the defect creation with the accumulated dose of positive ions. Up to -60 V, the dose and the energy of the positive ions are not high enough to have created defects noticeably changing surface state. Consequently, the ionization probability is probably still high. Above -60 V the signal is decreasing because the ionization probability is decreasing.

The experiments presented in Figure 11 and Figure 12 confirm previous conclusions made in this paper. The surface state of HOPG is probably similar for all biases, only the depth of the modified layer is changing. Indeed, due to the layer structure of HOPG, when positive ion energy is increased a new layer is reached and rapidly modified by impacting particles. The signal evolves fast to its steady state value. For MCBDD, the accumulation of impacts at increasing energy slowly destroy the hard structure of the diamond material and the signal slowly decreases to its steady state value.

### 3.5 Discussion

Based on experimental results (Figure 6) it can be noticed that diamond is the best among both carbon materials to promote negative-ion surface production at low positive-ion energy, from ~10eV/nucleon (-20V bias) to ~25eV/nucleon (-60V). The highest NI yield, observed here for diamond surface with low content of defects is believed to be due to diamond electronic properties which are expected to favour electron capture by incident particle and limits electron loss from the newly created negative-ion as discussed in reference 32.

All the results presented here have been obtained in deuterium plasma. However, generally speaking, no major difference between hydrogen and deuterium plasmas have been observed on what concerns negative-ion surface production on carbon materials (see for instance H$^-$ and D$^-$ yields versus surface temperature in H$_2$ [43] and D$_2$ [32] at -130V). In reference 31 a comparison between H$^-$ production on graphite in H$_2$ plasma and D$^-$ production in $D_2$ plasmas has been made. The main isotopic effect was observed on NIEDF shape. The $H^-$ NIEDF extends to higher energy than the D

.+0⁻ one. This was attributed to the higher energy deposited on the surface by positive deuterium ions during backscattering due to their higher mass. As a consequence, deuterium is expected to create more defects, or to create defects at slightly lower energy compared to hydrogen. Despite a complete detailed study of $H^-$ yield versus positive ion energy has not been undertaken, the global behaviour of $D^-$ yield in $D_2$ plasma is also observed for $H^-$ in $H_2$ (increase of yield with positive ion energy for graphite, decrease for diamond). Therefore, general trends are identical in $H_2$ and $D_2$ plasmas but absolute yields, threshold for sputtering onset, threshold for defect creation, or more generally speaking interaction with surfaces might change (see for instance comparison between deuterium and hydrogen operation of high density negative-ion sources[60]).

Unfortunately, no absolute yield measurement is obtained with the mass spectrometer measurement. However, recent measurements[53] show that mass spectrometer measurements could be put on an absolute scale by using a Magnetized Retarding Field Energy Analyzer[54]. This will be the subject of future works. In the meantime, there is only one direct comparison[55] in identical experimental conditions between carbon layers and cesiated surfaces which are routinely employed in high efficiency negative-ion sources. In this work, negative ion densities have been measured in a plasma where a diamond surface is introduced and biased between +20V and -30V. These densities have been compared to the densities obtained in the same plasma with no diamond surface (reference case) or with a caesiated stainless steel surface. The introduction of the caesitated surface leads an increase of negative ion densities in the plasma bulk by a factor 2.5 compared to the reference case whereas in presence of diamond surfaces no enhancement of negative ion density is observed. We have compared by the past relative NI production yields of graphite and diamond with those of usual metals (molybdenum, tungsten, stainless steel…)[56,57]. Carbon materials provides NI yield which are one to two orders of magnitude higher than metals. Hence our measurements are not in agreement with those of reference 55. As explained by the authors of reference 55 "the different plasma conditions and applied diagnostics limits the possibility for a direct comparison" between our works. In particular, we are working at much lower positive ion flux. We have shown that negative-ion surface production flux is proportional to the impinging positive ion flux[5] but the highest positive-ion flux reached in this experiment was still one order of magnitude lower than the one of reference 55. Enhanced erosion and diamond surface modifications at high positive ion flux might explain differences between both studies. Also, the experimental method used in the present study allows for a direct measurement of NI coming from the surface and is not affected by NI loss in the plasma volume which leads to a higher sensitivity of our technic to surface production.

Taking into account the exponential dependence of the ionization probability with the work function[58], it can be extrapolated from Wada measurements[59] at bias of -100 V that H- yield in hydrogen discharge increases by two orders of magnitude when going from pure molybdenum surfaces (work function ~ 4.5 eV) to cesiated low work function (< 2 eV) material. Therefore, carbon materials, which show yields one or two orders of magnitude higher than metals, would sit between clean metals and cesiated surfaces in terms of NI yield, at least under high energy positive ion bombardment (few tens of eV). However ceisated surfaces have proven to be efficient also at low energy particle bombardement (few eV), have demonstrated very low number of co-extracted electrons[60], and have the advantage to be continuously renewed by Cs evaporation while it might be difficult to keep a defect free diamond layer under plasma exposure due to its inherent reactivity with hydrogen, even at low energy[48].

# Conclusion

This work focuses on the production of negative-ions on graphite and diamond surfaces under deuterium plasma positive ion bombardment (10-60 eV/nucleon). First, it is shown that negative-ions can be efficiently self-extracted from the plasma even with a small voltage difference between the sample and the mass spectrometer orifice of only -10 V. Under this bias condition, the positive ion impact energy is on the order of 10 eV/nucleon.
The effect of the increase of positive ion energy on negative-ion production has been studied next. The negative-ion energy distribution functions (NIEDFs) have similar shapes for diamond and graphite whatever the impact energy. This suggests similar NI production mechanisms, with identical contribution of both sputtering and backscattering processes. However, the NI production yields, defined as the flux of negative-ions divided by the flux of positive-ions, behave completely differently with impact energy increase. The negative-ion production yield is decreasing for diamond surfaces, while it is strongly increasing for graphite. This increase is attributed to the onset of the sputtering mechanisms between 20 and 40 eV. The same mechanism occurs for diamond but is mitigated by a strong decrease of the ionization probability due to defect creation and loss of diamond electronic properties upon positive-ion impacts. As a consequence, the highest NI yield is observed for diamond surface with low content of defects. One can conclude that the electronic properties of pristine diamond are favourable for surface ionization of incident hydrogen particles. The present study suggests that electronic properties of insulators could be potentially used to promote surface production of $H^-/D^-$ negative ions.

# Acknowledgments

This work has been carried out within the framework of the French Federation for Magnetic Fusion Studies (FR-FCM) and of the Eurofusion consortium, and has received funding from the Euratom research and training programme 2014-2018 and 2019-2020 under grant agreement No 633053. The views and opinions expressed herein do not necessarily reflect those of the European Commission. Financial support was received from the French Research Agency (ANR) under grant 13-BS09-0017 H INDEX TRIPLED.